\documentstyle[prl,aps,epsfig,twocolumn]{revtex}

\draft

\newlength{\textwidthm}
\begin{document}
\title{High-temperature superfluidity of fermionic
atoms in optical lattices}
\author{W.~Hofstetter$^1$, J.~I.~Cirac$^{1,2}$, P.~Zoller$^{1,3}$, E.~Demler$^1$,
and M.~D.~Lukin$^1$}
\address{$^1$ Physics Department, Harvard University, Cambridge, MA 02138 \\
$^2$ Max-Planck Institute for Quantum Optics, Garching, Germany \\
 $^3$ Institute for Theoretical Physics, University of Innsbruck, Austria}
\date{\today}
\maketitle

{\bf The experimental realizations of degenerate Bose
\cite{boserev} and Fermi \cite{fer1,fer2,fer3,fer4} atomic samples have
stimulated a new wave of studies of quantum many-body  systems in
the dilute and weakly interacting regime. The intriguing
prospective of extending these studies into the domain of strongly
correlated phenomena is hindered by the apparent relative weakness
of atomic interactions.  The effects due to interactions can,
however, be enhanced if the atoms are confined in optical
potentials created by standing light waves \cite{boseth,bloch,kasevich}.
The present Letter shows that these techniques, when applied to
ensembles of cold fermionic atoms, can be used to dramatically
increase the transition temperature to a superfluid state and thus
make it readily observable under current experimental conditions.
Depending upon carefully controlled parameters, a transition to a
superfluid state of Cooper pairs \cite{tinkham}, antiferromagnetic
states \cite{o} or more exotic d-wave pairing states \cite{dwave}
can be induced and probed. The results of proposed experiments
can provide a critical insight into the origin of high-temperature
superconductivity in cuprates \cite{hihgtc}. }

An active search is now under way to implement a BCS transition of
degenerate fermionic gases to a superfluid state analogous to
superconductivity \cite{stoof1,stoof2}. However, in free space or weakly
confining atom traps the transition temperature to the superfluid
state scales exponentially with interaction strength, $k_B
T_c^{\rm free} \approx 0.3 E_F^{\rm free} \exp[-\pi/(2 k_F
|a_s|)]$, with $E_F^{\rm free}$ the Fermi energy. For a dilute
atomic gas the product of Fermi momentum and scattering length
$k_F |a_s| <<1$, which makes the transition temperature
exceedingly low.
%The efforts are presently being
%directed toward increasing the atomic interaction strength by means
%of the so-called Feshbach resonances. However departures from the dilute
%regime comes at price of enhanced losses which have particularly sevear
%consqeunces for Fermionic systems.

Atoms in potentials created by standing light waves (optical
lattices) tend to localize near potential nodes thereby increasing
the strength of effective interactions, whereas the kinetic energy
provided by tunneling between the different sites can be strongly
suppressed \cite{boseth}. Very recently, fascinating experiments
involving bosonic atoms in optical lattices \cite{bloch} revealed
a quantum phase transition from a superfluid to Mott insulating
state \cite{boseth,fisher}. Fermionic atoms confined in an optical
lattice can undergo a phase transition to a superfluid state at a
temperature that exceeds that of weakly confined atoms  by several
orders of magnitude. Attractive atomic interactions result in an
s-wave pairing in which case fermionic atoms can undergo a
BCS-type transition. In what perhaps is the even more intriguing
prospective, fermionic atoms with repulsive interactions
correspond to an experimental realization of a Hubbard model that
is widely discussed for strongly correlated electron systems such  as
high-$T_c$ cuprates \cite{A}. In particular, d-wave superconducting states
have been
conjectured to exist in such systems, but so far this conjecture
eluded rigorous confirmation. We show that atomic systems with
carefully controllable parameters and a variety of precise tools
to detect the resulting phases can be used  to provide a critical
insight into this outstanding problem. In essence, this approach
can be viewed as an implementation of the pioneering ideas due to
Feynman \cite{fynman} for simulations of one quantum system by
another.

Consider an ensemble of fermionic atoms illuminated by several
orthogonal, standing wave laser fields tuned far from atomic
resonance. These fields produce a periodic potential for atomic
motion in two (or three) dimensions of the form $V(x) = V_0
\sum_{i=1}^{2(3)} {\rm cos}^2(k x_i)$ with $k$ the wave-vector of
the light. The potential depth $V_0$ is typically expressed in the
units  of the atomic recoil energy $E_R = \hbar^2 k^2/2m$. We will
be interested in the situation in which there is roughly one atom
per lattice site. Such atomic densities correspond to free-space
Fermi energies on the order of $E_F^{\rm free} = (3/\pi)^{2/3}
E_R$.
%??This process is either assisted or inhibited
%by interactions.
At the same time the atoms can tunnel from one site to another. We
assume that two kinds of atoms are present, differing by angular
momentum or generalized spin ($\sigma = \{\uparrow,\downarrow\}$).
For sufficiently low temperatures the atoms will be confined to
the lowest Bloch band, and the system can be described by a
Hubbard Hamiltonian \cite{boseth,A}
\[
H = - t
\sum_{\{i,j\},\sigma} (c_{i,\sigma}^+c_{j,\sigma} +
c_{j,\sigma}^+c_{i,\sigma}) + U \sum_i n_{i,\uparrow}
n_{i,\downarrow},
\]
where $c_{i,\sigma}$ are fermionic annihilation operators for
localized atom states of spin $\sigma$ on site $i$,
$n_{i,\sigma} = c^+_{i,\sigma} c_{i,\sigma}$. We
assume that the average occupation numbers $\langle
n_{i,\uparrow;\downarrow} \rangle \le 1$. The parameter $t$
corresponds to the tunneling matrix element between adjacent
sites, $t = E_R (2/\sqrt{\pi}) \xi^3 \exp(-2\xi^2)$, and  the
parameter $U = E_R a_s k \sqrt{8/\pi} \,\xi^3$ characterizes the
strength of the on-site interaction with  $\xi = (V_0/E_R)^{1/4}$.
The sign of the scattering length $a_s$ determines the nature of
atomic interactions: negative $a_s$ corresponds to attraction
between atoms, whereas positive $a_s$ corresponds to repulsion.
For example, in the case of Li$^6$ \cite{fer2,fer3} both cases can be
realized depending on the particular electronic states that are
being trapped in a lattice.
%For example, electron-spin polarized Li$^6$ atoms in the
%states with different orientation  of nuclear spin
%$|m_s =1/2,m_i=0\rangle$ and
%$|m_s=1/2,m_i = +1 \rangle$ attract each other,
%whereas the scattering length associated with the states
%$|m_s = -1/2,m_i = 0\rangle$ and $|m_s=-1/2,m_i = 1 \rangle$
%can be  varied from positive to negative by small changes
%of a magnetic field across the so-called Feshbach resonances.

%{\it Attractive phases.}
Consider first the situation corresponding to negative $U$. The effect of the lattice on the superfluid
transition can be best understood starting from the limit of large tunneling $t \gg |U|$. Here, similar to the
free-space transition,  the interaction strength is much weaker than the kinetic energy and the ground state of
the system is then given by a ``standard'' BCS wave function, with an energy gap and transition temperature
$T_c$ that depend upon $t$ and $u$ \cite{swave}. For the lattice with filling fraction near unity ($\langle
n_{\uparrow} \rangle + \langle n_{\downarrow} \rangle \sim 1$) the Fermi energy is on the order of $t$, the
density of states scales as $N/t$ (with $N$ being the total number of lattice sites)  and the average two-atom
interaction strength as $U/N$. Standard BCS theory can be applied to predict a critical temperature $T_c$ that
for a 3-D situation scales as $k_B T_c \approx 6  t \exp(- 7 t/|U|)$ 
\cite{swave}. An increase in the depth of the
optical potential results in stronger atom localization and hence an increased interaction strength $U$. At the
same time, the tunneling $t$ becomes weaker. The combined effect of these two factors is a dramatic increase in
$T_c$ due the exponential factor that competes with a very modest linear reduction due to a decreased Fermi
energy.

As the tunneling becomes comparable to the on-site interaction, the BCS picture is no longer valid. Due to
strong attraction, atoms form pairs within single lattice sites. The entire system can then be considered as
an ensemble of composite bosons. They can tunnel together at a rate $\sim t^2/|U|$, by virtual transitions via
intermediate singly occupied states. In this regime non-ordered pairs exist at high temperatures, whereas the
superfluid state - a condensate of composite bosons - appears below $k_B T_c \sim t^2/|U|$.  Clearly, in this
limit the increase in the potential depth will lead to a reduced mobility of pairs and hence a decrease in
$T_c$. The maximal critical temperature $T_c^{\rm max}$ is  achieved at the crossover between the two regimes,
when interaction and tunneling are comparable (more precisely at $U \sim 10 t$), which corresponds to a
potential depth $\xi^2 \approx 1/2 {\rm log}[5 \sqrt{2}/k |a_s|]$ and
\begin{equation}
k_BT_c^{\rm max} \approx 0.3 E_F^{\rm free} k |a_s|.
\end{equation}
That is, the critical temperature for atomic fermions trapped in a lattice scales only {\em linearly} with the
small parameter $k |a_s|$. This accurate result for the critical temperature is based on nonperturbative Monte-Carlo simulations of the 
fermionic Hubbard model
\cite{swave}.

Several specific approaches to achieve the superfluid state can be considered. For example, the optical
potential can be adiabatically turned on, starting from a weakly confined Fermi-degenerate mixture of the two
atomic states of appropriately chosen density. In this procedure the atomic quasi-momentum is approximately
conserved but the band-structure associated with the periodic potential changes, resulting in a non-equilibrium
distribution, with an effective temperature $T_f$ different from the initial $T_{\rm in}$. The final temperature
$T_f$ can easily be estimated from the relation $ \sum_{\bf k} \epsilon_{\bf k} f(\epsilon^0_{\bf k}/T_{in}) =
\sum_{\bf k} \epsilon_{\bf k} f(\epsilon_{\bf k}/T_f) \label{Tf} $, where $f(x)=1/(e^x+1)$ is the Fermi-Dirac
distribution function, $\epsilon^0_{\bf k} =k^2/2m-\mu^0$ is the original dispersion of atoms in free space, and
$ \epsilon_{\bf k} = - 2 t ( {\rm cos}(k_x a) + {\rm cos}(k_y a) + {\rm cos}(k_z a))-\mu$ is the dispersion in a
tight--binding model with $a=\pi/\lambda$ the lattice period. Two important processes determine $T_f$. First of
all, the presence of the lattice makes the system anisotropic and hence changes the shape of the Fermi surface.
This results in an effective heating of the system as $V_0$ increases from zero to about $E_R$. As $V_0$ is
increased even further the shape of the Fermi surface remains approximately the same and only the Fermi velocity
(or effective atomic mass) changes leading to a reduction of the effective temperature within the lowest Bloch
band (see Fig.~1). This suggests that it is optimal to turn on a weak lattice potential while the fermionic
sample is in contact with a cooling reservoir (e.g. atomic BEC). Although the resulting phase space density will
not be significantly altered, this loading procedure will allow to avoid the heating which is present at the
initial stages of creating an optical lattice. When $V_0\sim E_R$ (point C in Fig.~1a) the system is decoupled
from the reservoir and the lattice potential is increased until the transition to the superfluid phase is
reached.

To control precisely the resulting quantum phase (especially in the situations described below) an accurate
manipulation of the filling fraction may be important. This can be achieved, for example, if atoms with three
internal states are used. If a dense, degenerate ensemble is prepared in a state that is not affected by the
optical lattice, a laser driven Raman transition into a pair of trapped spin states can be used to produce
exactly one atom (or its fraction) per each lattice site. The essential idea of this approach is that the
energy shifts associated with atom interactions and the Pauli principle can be used to block the transitions
into states with more than one atom per lattice site. In this case an effective $\nu \pi$-pulse will result in
a filling fraction of $\nu$ with an uncertainty that scales at most with the inverse size of the lattice.

Let us now consider the implications of these results in the light of present experimental possibilities. The
relevant temperature calculated numerically for Li$^6$ \cite{fer2} and K$^{40}$ \cite{fer1} atoms is shown in
Fig.~1. In this figure we consider a Li$^6$ atomic sample of a very modest density corresponding to a unity
filling in an optical lattice produced by CO$_2$ laser ($\lambda = 10\mu m$). For such densities, a dramatic
increase in the critical temperature is possible. Note in particular that a phase transition can
be achieved
starting from an initial temperature of about $0.1 E_f^{\rm free}$. The
CO$_2$ lattice has the additional advantage of exceptionally long lifetimes, which should be sufficient to
achieve the transition even for relatively low energy scales involved. Another  scenario is to trap Li atoms
in an optical lattice created by a Nd:YAG laser $\lambda\sim 1.06 \mu m$. Although in this case the densities
of  $10^{12}-10^{13}$ cm$^{-3}$ will correspond  to a filling fraction slightly less than unity, the resulting
critical temperature can still be in the range of $0.1 E_F^{\rm free}$.  The inset shows a diagram for K$^{40}$ atoms
trapped in a similar lattice.
As indicated by the two cases presented in Fig.~1 the transition
to a superfluid state is expected to occur for the same initial temperature
if cooling due to adiabatic switching on of the lattice is taken into account.
Therefore, the value of the maximal initial temperature
at which a phase transition can occur
is {\it almost independent of the scattering length} and corresponds
to about one tenth of the free space Fermi energy.

Before proceeding we point out that in contrast to the approaches that are based on increasing the scattering
length or atom densities, which result in a very interesting regime of BCS--BEC crossover
\cite{timm,holland,ohashi}, an optical lattice does not lead to an enhancement of inelastic loss processes but
rather their suppression, as no more than two atoms can ever occupy the same lattice site. We also note that
this implies an extremely low spin flip rate between different atomic states, thereby allowing comparatively
large magnetic fields to tune the scattering lengths without significantly perturbing the relative fractions of
$n_\downarrow$ and $n_\uparrow$. It is also important to point out that although the free-space ensembles at
densities $n |a_s|^3 > 1$ can no longer be considered as weakly interacting gas \cite{timm,holland}, the Hubbard
model nevertheless remains a valid description for the lattice case even in the regime of very strong
confinment.  Here corrections to the on--site interaction can be accurately derived from known molecular
potentials. Finally we emphasize again, that in contrast to a weakly confined Fermi-gas,  the critical
temperature for the optical lattice filled with attractive atoms (Fig.~1) can be  predicted very accurately even
in the most interesting, intermediate, regime $t\sim |U|$, since the behaviour of the Hubbard model for this case is by now very well understood.

It is intriguing to consider possible extensions of the above
ideas to a situation in which different atoms repel each other
($a_s>0$). This is realized for $|\downarrow\rangle = |F
=9/2,m_F=9/2\rangle$ and $|\uparrow\rangle =  |F=9/2,m_F = 7/2
\rangle$ states  of atomic K at zero magnetic field \cite{fer1}.
In this case it is energetically unfavorable for two atoms to be
on the same lattice site. However, adjacent atoms can virtually
tunnel to the same site. This process lowers the
total energy of two atoms in adjacent wells, thereby creating
effective hard-core attractive interactions of different spins.
When the filling fraction of the lattice is close to one, this
leads to a ground state in which adjacent sites are
always occupied by atoms with alternating spins (see Fig.~2), i.e.~an
antiferromagnetic phase.

What happens when the filling fraction is
smaller than one? This question lies at the heart of the present
debates on the nature of high-temperature superconducting
cuprates.
%Repulsive Hubbard model is believed to provide a reasonable description
Clearly, a strong on-site repulsion makes it unfavorable for atoms
to bind into the usual s-wave pairs, in which the probability to
be at the center of orbit should be  maximal.  However, it has
been conjectured \cite{dwave} that anisotropic d-wave
pairs can be formed, which can result in a d-wave superfluid
phase capable of explaining many of the observed properties in
cuprates.

$d$-wave Cooper pairs can be thought of as spin singlet angular momentum $l=2$ Cooper pairs. In the absence of
the lattice potential all five $l_z=\pm 2$, $\pm 1$, $0$ components are degenerate. A cubic potential splits
these states according to the representations of the cubic point group, and  we find triply and doubly
degenerate Cooper pairs \cite{sigrist}. When the symmetry of the crystal is lowered even more, as e.g. in a
two dimensional lattice, the degeneracies between various Copper pairs are removed further, and in the $x$,$y$
directions the lowest energy Cooper pair is described by the momentum space wave function \cite{Leggett}
$\Delta({\bf p})= {\rm cos}(p_xa)-{\rm cos}(p_ya)$. Although so far the existence of d-wave superconductivity
in the repulsive Hubbard model eluded rigorous confirmation, we believe that these ideas can now be tested
experimentally in ensembles of fermionic atoms. For example, Fig.~2 shows a phase diagram for the system
of repulsive atoms in two dimensions \cite{dwave} calculated using the FLEX approximation \cite{flex}.
Although the resulting $T_c$ is believed to be somewhat lower
than in the s--wave case, this
calculation suggests the existence of the d--wave phase for
feasible atomic temperatures and densities.

We next consider several approaches that can be used  to detect
and accurately probe the  resulting quantum phases. For example,
interference of the atoms released from the lattice has been used
to probe the superfluidity of bosons \cite{bloch}. Due to the exclusion
principle, identical fermions can never be in the same quantum
state. This implies that in the degenerate regime atomic interference
patterns due to each momentum state will be superimposed,
resulting in real space interference peaks which reflect
the shape of the Fermi surface.
With the appearance of pairing, the atomic momentum in the pairs becomes on
the order of Planck's constant divided by the size of the pair. As
a result, the momentum distribution is no longer a sharp step
function, which will be directly reflected in the interference
pattern as shown in Fig.~3.

In order to directly detect superfluidity of the pairs,
photoassociation spectroscopy \cite{association} can be used.
Weakly bound Cooper pairs can be converted into molecules
by using a laser--induced transition into a bound molecular state.
The interference pattern of the released
bosonic molecules will then provide extremely sharp peaks due to the
presence of a superfluid fraction,
in direct analogy to ref. \cite{bloch}.

The spectrum of elementary excitations also provides an accurate
probe for the nature of the quantum phase. It can be measured in a
system of cold atoms by exciting the motional states of atoms
using laser pulses. For example, atoms can experience Bragg scattering
off two non-collinear laser beams, provided that the frequency
difference $\delta\omega$ of the lasers matches the resonance frequency of
elementary excitation with momentum $q$ determined  by the angle
between two lasers (Fig.~4) \cite{bragg}. This technique
provides a direct measurement of the density--density correlation
function.
By monitoring the number of Bragg--scattered atoms as a function
of $\delta\omega$, the presence of a superfluid phase can be
detected unambiguously:
One observes a sharp peak due to a collective Bogoliubov mode,
which is separated from a broad feature (corresponding to quasi--particle
excitations) by an energy gap (see Fig.~4).
In the normal state, on the other hand, the continuum excitations
are gapless and the collective mode is not visible due to strong damping.

With this technique, it is also possible to detect d--wave superfluidity.
For example, Fig.~5 shows the onset frequency of the quasiparticle
continuum corresponding to the d--wave superfluid phase.
A strong anisotropy, together with a vanishing gap for certain
momenta can provide unambiguous evidence for the presence of such
a phase.

It should be noted that in addition to the effects described here,
a number of other intriguing avenues can be considered. Atomic systems
make it possible to accurately probe the effects of dimensionality.
Varying for example the depth of the trapping potential
in one direction while keeping the other parameters fixed,
it should be possible to observe a transition between a two-dimensional and a
three-dimensional situation.
One can also envision a situation in which more than two
internal atomic states are trapped in the optical lattice. Such a
system is expected to result in a variety of new quantum phases
such as superconductivity based on quadruplets rather than pairs.
Likewise, these systems might allow for accurate studies of the effect
of dissipation and decoherence on macroscopic quantum
phases. Another interesting phenomenon that may be studied 
by creating a system with unequal densities for the two spin states
is the possibility  of the Fulde-Ferrel-Larkin-Ovchinnkov
(FFLO) state \cite{LO,FF} with a modulated superfluid order
parameter.

\def\etal{\textit{et al.}}

%\newpage

%%%%%%%%%%%%%%%%%%%%%%%%%%%%%%%%%%%%%%%%%%%%%%%%%%%%%%%%%%%%%%%%%%%%%%%%%%%%

\begin{figure}[ht]
\begin{minipage}{0.4\linewidth}
\begin{center}
\epsfig{file=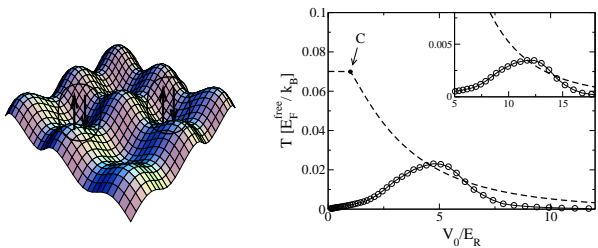,width=0.99\linewidth}
\end{center}
\end{minipage}
\hfill
\begin{minipage}{0.55\linewidth}
\begin{center}
\epsfig{file=fig2.eps,width=0.9\linewidth}
\end{center}
\end{minipage}
 \vspace*{2ex}
 \caption{ Left: Attractive fermionic atoms in optical lattices can undergo
pairing resulting in a BCS transition into a superfluid state. Right: Critical temperature for a transition of
Li$^6$ atoms (circles) into the superfluid state as a function of the optical lattice depth in a
three--dimensional CO$_2$ lattice. Li atoms in the states with different orientation  of nuclear spin
$|\downarrow\rangle = |F = 1/2,m_F = -1/2\rangle$ and 
$|\uparrow\rangle =  |F = 1/2,m_F = 1/2 \rangle$ are considered
at a magnetic field of $\sim 0.1$ T which corresponds to a scattering length of $a_s \sim -2.5 \times 10^3 a_0$
in atomic units. The absolute energy scale is given by $t/\hbar \approx 0.5 {\rm kHz}$ at the phase transition.
For the same values of the scattering lengths and densities the free-space BCS formula gives for the temperature
of the superfluid transition $T_c^0= 1.6 \times 10^{-12} E_F^{\rm free}/k_B$ for Li$^6$. The inset shows the
analogous plot for $K^{40}$  atoms in a Nd:YAG lattice at half filling. For K$^{40}$ we took two states
characterized by total angular momentum $|\downarrow\rangle = |F =9/2,m_F=-9/2\rangle$ and $|\uparrow\rangle =
|F=9/2,m_F = -7/2 \rangle$ at a magnetic field above a Feshbach resonance and choose $a_s \sim -2.\times 10^2
a_0$. As discussed in the text, turning on an optical lattice also changes the temperature of the system. It is
most advantageous to turn on a weak lattice potential $V_0/E_R \sim 1$ while the atoms are cooled. The cooling
is then switched off (point C in the figure) and the lattice depth is increased adiabatically, reducing the
effective temperature (shown by the dashed line for an initial temperature $T_{\rm in}=0.07 E_F^{\rm
free}/k_B$). }
\end{figure}

%%%%%%%%%%%%%%%%%%%%%%%%%%%%%%%%%%%%%%%%%%%%%%%%%%%%%%%%%%%%%%%%%%%%%%%%%%%%

\begin{figure}[h]
\begin{minipage}{0.4\linewidth}
\begin{center}
\epsfig{file=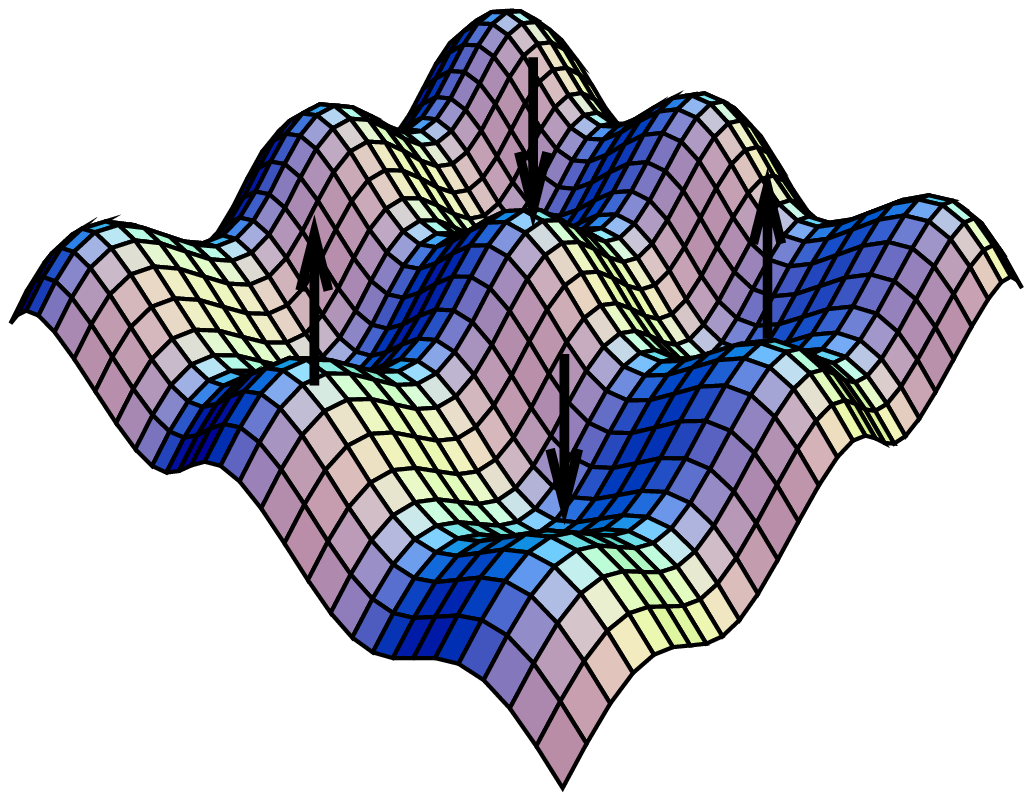,width=0.99\linewidth} \\[0.5cm]
\epsfig{file=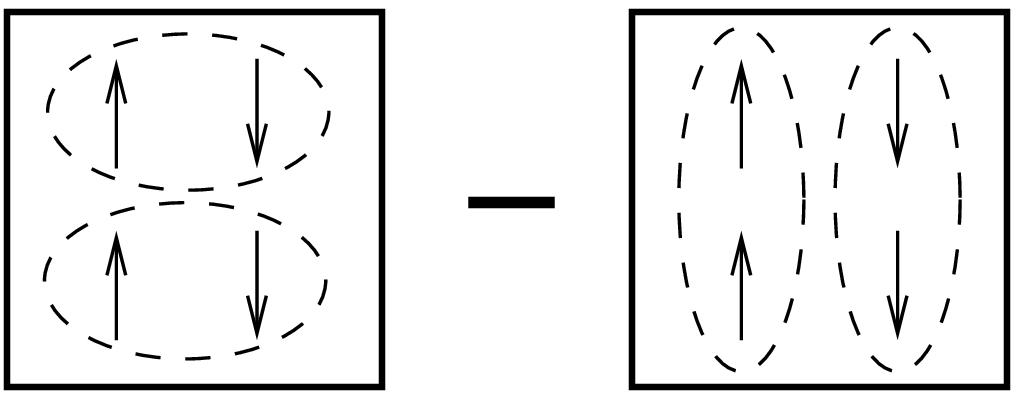,width=\linewidth}
\end{center}
\end{minipage}
\hfill
\begin{minipage}{0.55\linewidth}
\begin{center}
\epsfig{file=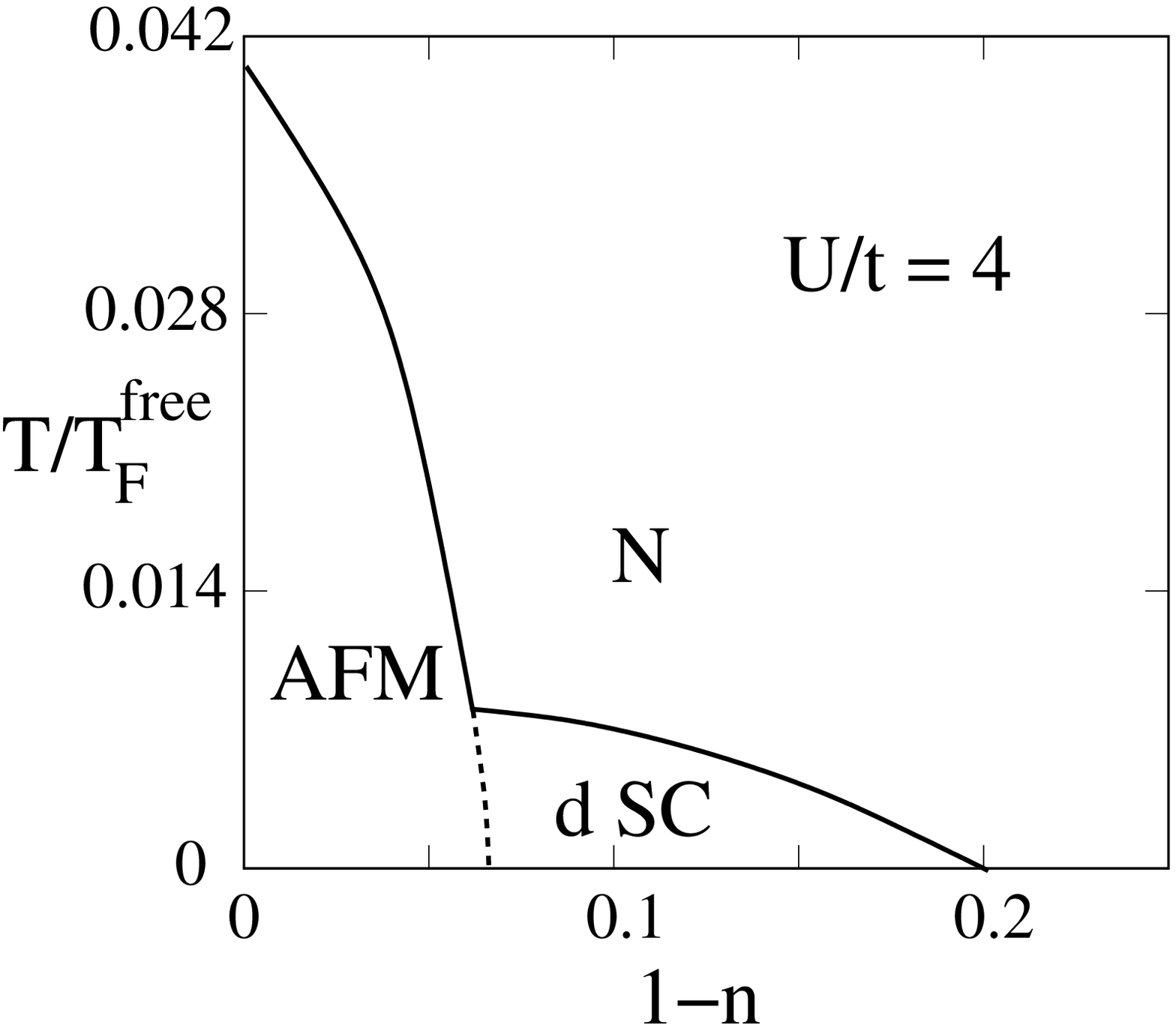,width=0.9\linewidth}
\end{center}
\end{minipage}
 \vspace*{2ex}
 \caption{In the case of repulsion
it is energetically unfavorable for two atoms to be the on the
same lattice site. However, virtual tunneling creates effective
hard--core attractive interactions of different spins on nearby
sites. This may result either in antiferromagnetic (upper left) or
d-wave superfluid phases, depending upon the filling fraction $n$.
In three dimensions and at
filling fraction $n=1$ the maximum critical temperature for
antiferromagnetic ordering is identical to the superfluid case,
and of the order $0.1E_f^{\rm free}$.
The d--wave symmetry of the superfluid is related to the
``parent'' insulating phase (see lower left).
The phase diagram for repelling Li$^6$ atoms in a 2D lattice,
calculated using the FLEX approximation, is shown on the right hand
side. Adiabatic cooling due to switching on of the lattice
has been taken into account.
The situation we have in mind corresponds to a very large potential
depth along one direction and an equal, finite depth along the other
two directions. This results in a set of weakly coupled 2D lattices.
Within each lattice tunneling is given by $t$.
We note that the repulsive Hubbard model may also have phase
separation for some filling factors, which corresponds to
immiscibility of the two spin species of the atoms.
%For a detailed
%discussion of the arguments in favor and against phase separation
%in the positive Hubbard model the reader is referred to Ref...
}
\end{figure}

\begin{figure}[h]
\begin{minipage}{0.45\linewidth}
\begin{center}
\epsfig{file=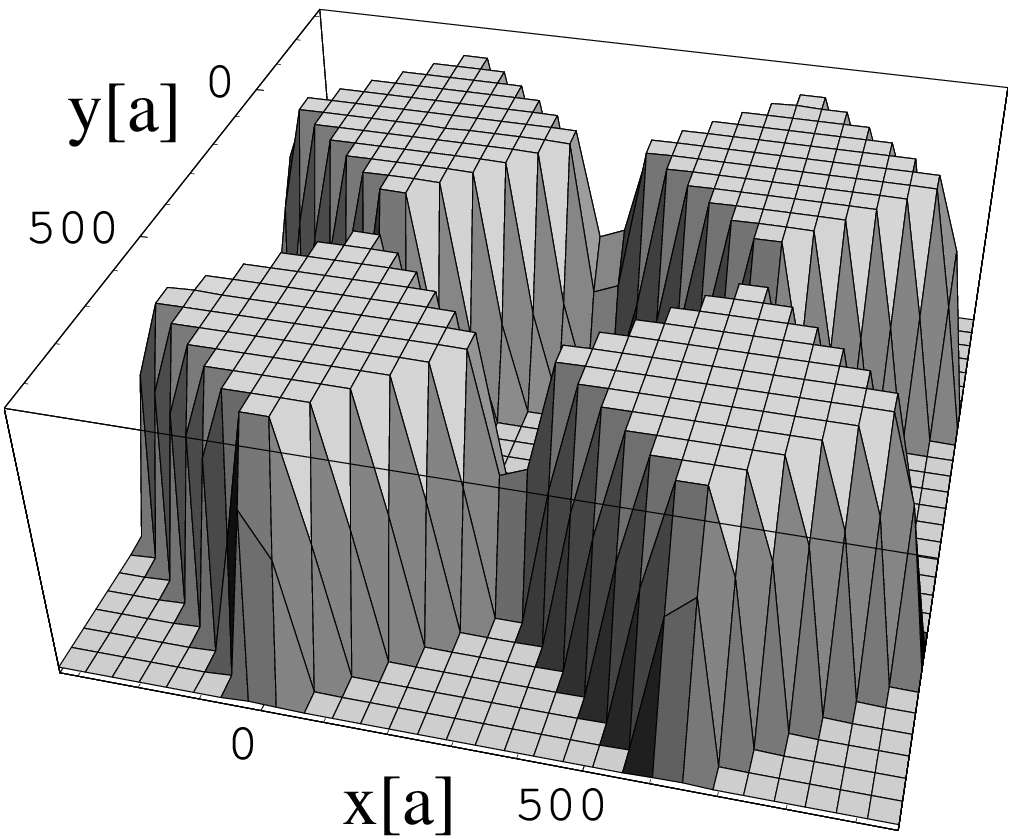,width=\linewidth}
\end{center}
\end{minipage}
\hfill
\begin{minipage}{0.45\linewidth}
\begin{center}
\epsfig{file=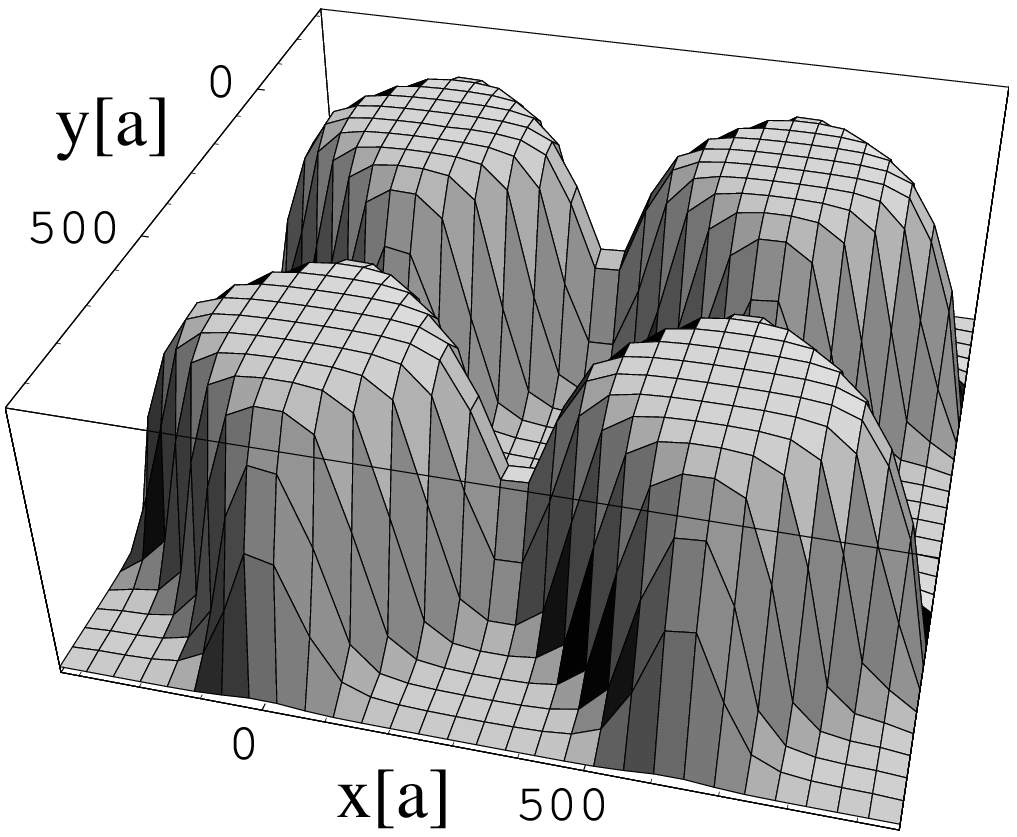,width=\linewidth}
\end{center}
\end{minipage}
 \vspace*{2ex}
 \caption{
Central part of the atomic interference pattern
after a free expansion for $t=500 \times (\hbar/E_R)$.
We are considering a 10x10 optical lattice
($a$ denotes the lattice constant).
Left: normal state at half filling. The sharp edges of the
interference peaks reflect the atomic momentum distribution:
in the normal state the peaks will become sharper with
decreasing temperature -- a direct consequence of Fermi-statistics.
Right: BCS state at half filling, with a gap $\Delta/t=0.6$
corresponding to an interaction strength $U/t \approx -2.5$.
}
\end{figure}

\begin{figure}[h]
\begin{minipage}{0.5\linewidth}
\begin{center}
\epsfig{file=fig8.eps,width=\linewidth}
\end{center}
\end{minipage}
\hfill
\begin{minipage}{0.4\linewidth}
\begin{center}
\epsfig{file=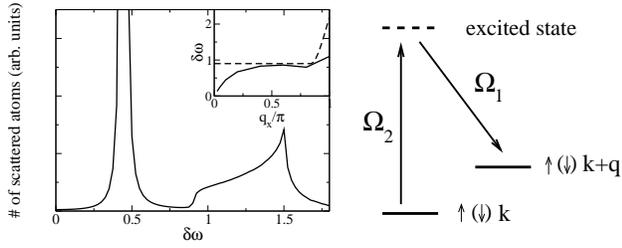,width=\linewidth}
\end{center}
\end{minipage}
 \vspace*{2ex}
 \caption{Left:
Bragg scattering of atoms off laser beams
with frequency difference $\delta\omega$ and
wave-vector difference $q_x=q_y=0.1\pi$ for attractive fermions at
filling $n=0.6$ and $U/t=-2.5$,
corresponding to an s--wave gap $\Delta/t \approx 0.45$.
The spectrum consists of a sharp collective mode
and a continuum at higher frequencies.
The sharp resonance corresponds to long--wave coherent
excitations of the condensate. The continuum arises
due to the breaking of Cooper pairs, which results
in quasiparticle excitations.
In the inset we show the dispersion of the collective mode
(solid line) and the onset frequency of the continuum (dashed line).
Right: schematic picture of the two--photon process involved.
$\Omega_1$ and $\Omega_2$ are the Rabi rates of the optical fields.
}
\end{figure}

\begin{figure}[h]
\begin{minipage}{0.4\linewidth}
\begin{center}
\epsfig{file=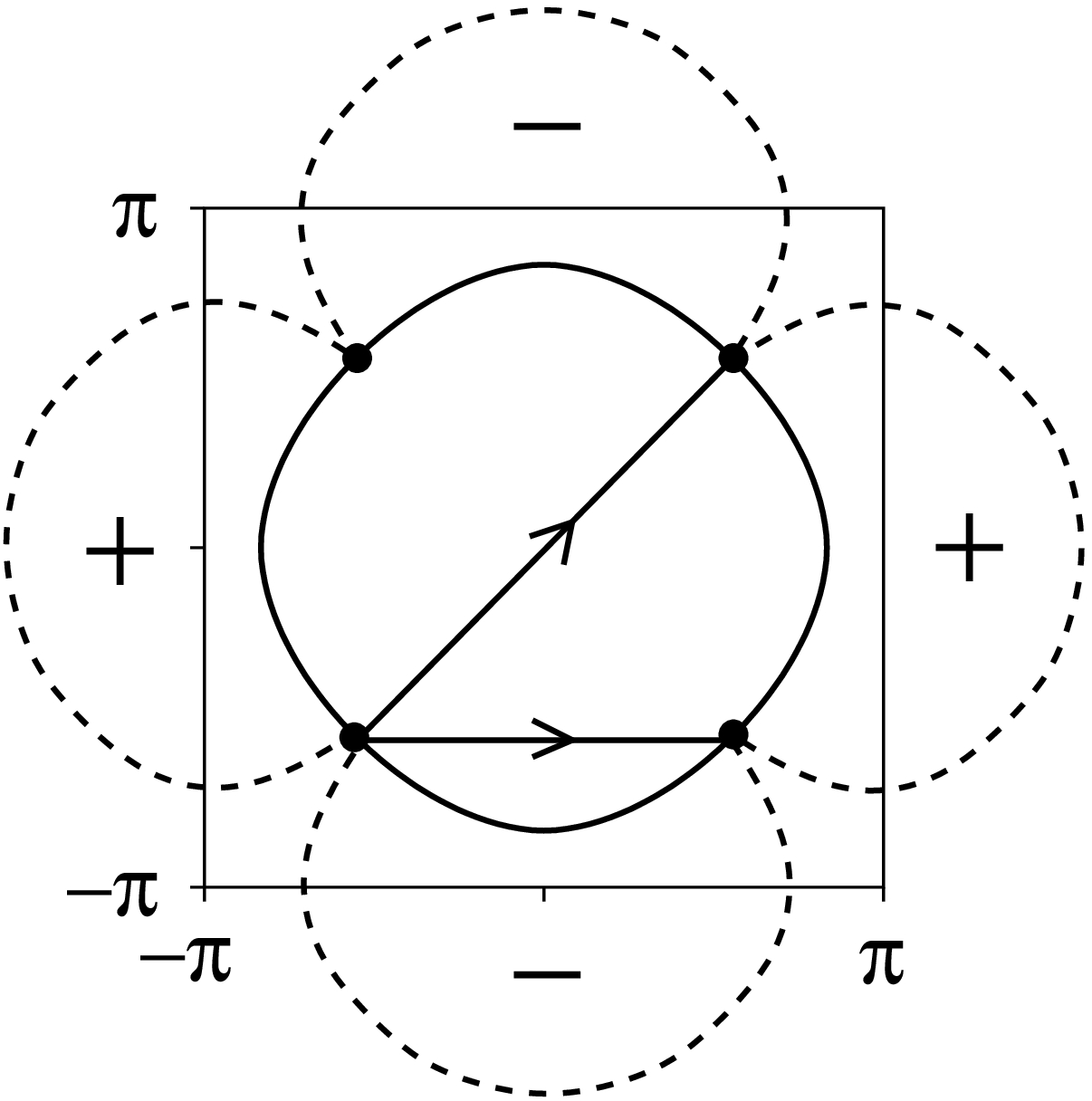,width=\linewidth}
\end{center}
\end{minipage}
\hfill
\begin{minipage}{0.5\linewidth}
\begin{center}
\epsfig{file=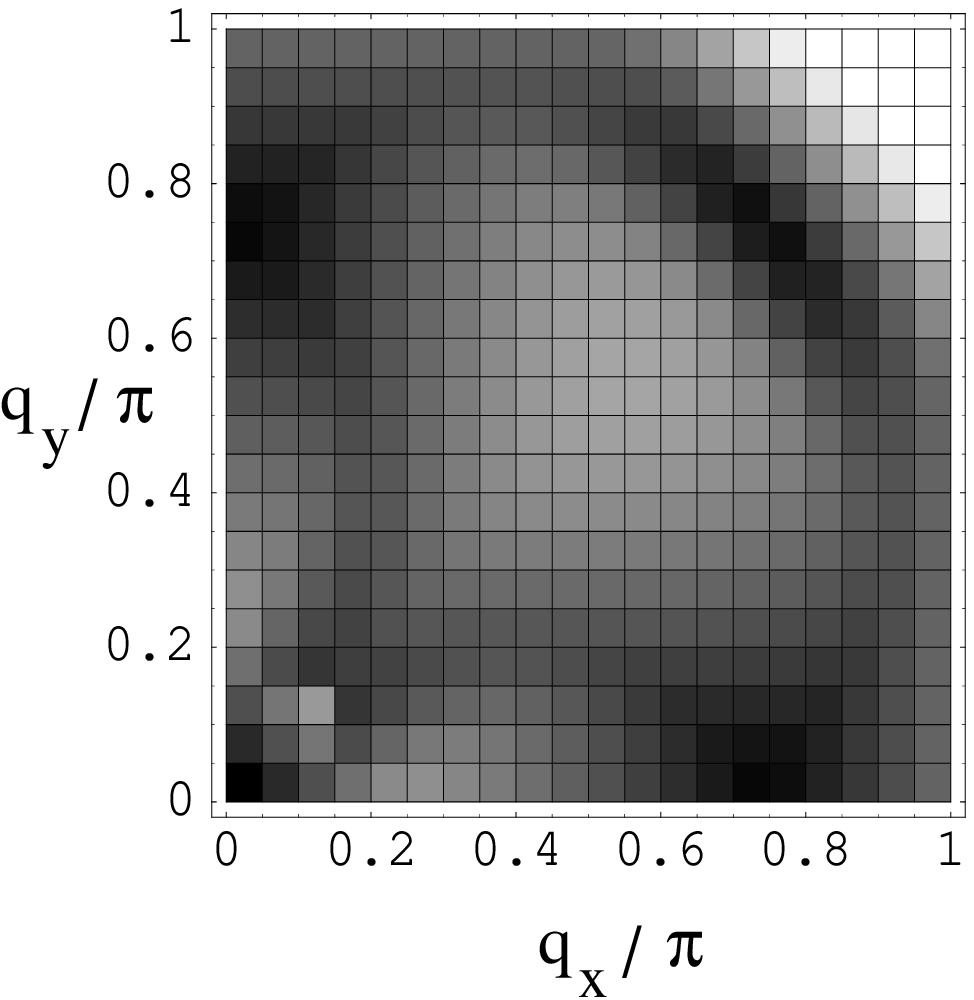,width=\linewidth}
\end{center}
\end{minipage}
 \vspace*{2ex}
 \caption{Probing d--wave pairing at filling $n<1$ via Bragg scattering.
Left: schematic diagram of the Fermi surface (solid line) and
the $q$--dependence of the
Cooper pair wave function $\Delta(q_x,q_y)$ (dashed line).
At the four nodal points shown by black dots,
the wave function and the quasiparticle excitation energy vanishes.
For the special wave vectors connecting
these points, the density response is gapless (black spots in the right figure).
Right: onset frequency
$\omega_{\rm min}(q_x,q_y)$ of the quasiparticle continuum,
dark regions corresponding to low frequencies (vanishing gap). }
\end{figure}

%%%%%%%%%%%%%%%%%%%%%%%%%%%%%%%%%%%%%%%%%%%%%%%%%%%%%%%%%%%%%%%%%%%%%%%%%%%%
\end{document}